# Itinerant-electron Ferromagnetism in $W(Nb)O_{3-\delta}$

I. Felner, U.Asaf and M. Weger

Racah Institute of Physics, The Hebrew University, Jerusalem 91904, Israel.

S. Reich and G. Leitus

Department on Materials and Interfaces, The Weizmann Institute of Science, Rehovot, 76100, Israel

The crystal structure and the magnetic properties of the $W_{1-x}Nb_xO_{3-\delta}$, (x<0.03) system have been investigated. In contrast to the orthorhombic diamagnetic $WO_3$, the material with x=0.01 is paramagnetic down to 5 K. Introducing of 2.5 at. % of Nb into $WO_3$ leads to a tetragonal structure and to a weak itinerant ferromagnetic ordering below $T_C$= 225 K. The saturation magnetic moment at 5 K is $1.07*10^{-3}$ $\mu_B$, whereas the paramagnetic effective moment is 0.06 $\mu_B$ per mole. This high ratio indicates itinerant ferromagnetism in $W_{0.975}Nb_{0.025}O_{3-\delta}$.

## Introduction

During the course with superconducting materials, our attention was drawn to the recent report of 2D [1] superconductivity (SC) at 91 K in $Na_{0.05}WO_3$. Stoichiometric $WO_3$ has an orthorhomic structure and it is diamagnetic insulator material, since the $W^{6+}$ 5d band is empty. When low concentration of $Na^{1+}$ ions are added, they donate their 3s electrons to the W 5d band, resulting in bulk metallic behavior and SC at 91 K. The heart of the present investigation is the assumption that it is possible to get SC in dilute substitution of $Nb^{5+}$ into $WO_3$ samples, despite the low $N(E_F)$ [2]. For this reason we have prepared the $W_{1-x}Nb_xO_3$ system with low amount of Nb concentration. A detailed study of the phase diagram [3] of the binary system $Nb_2O_5$-$WO_3$ has shown, that indeed, only 3 mol% of $Nb_2O_5$ can be dissolved in tungsten trioxide at high temperatures.

This paper aims the magnetic properties of the $Nb_2O_5$-$WO_3$ solid solution system in a narrow range of $Nb_2O_5$ concentrations. The magnetic susceptibility of stoichiometric $Nb_2O_5$ is temperature independent, but the oxygen deficient reduced $Nb_2O_{5-2x}$ forms, show an increasing paramagnetic behavior with x, [4,5]. On the other hand, the major component namely $WO_3$, is diamagnetic [6]. We show that $W_{0.99}Nb_{0.01}O_{3-\delta}$ is not SC down to 5 K and exhibits a typical paramagnetic behavior. Surprisingly, for the tetragonal $W_{0.975}Nb_{0.025}O_{3-\delta}$, a long range weak ferromagnetic signal is found below $T_C$=225 K. The same tetragonal structure is found at 120 K (below $T_C$), suggesting that the ferromagnetic transition is not connected with a crystallographic phase transition. The saturated magnetic moment at 5 K is $1.07*10^{-3}$ $\mu_B$ , whereas the paramagnetic effective moment $P_{eff}$ =0.06 $\mu_B$ per mole. The high ratio between the two values suggests that $W_{0.975}Nb_{0.025}O_{3-\delta}$ is an itinerant ferromagnet.

**Experimental details.**

Mixtures with concentrations 1.0 and 2.5 mol % of niobium pentoxide in $WO_3$ were pressed into pellets, which were packed and sealed in a platinum tube to avoid the sublimation of the $WO_3$. After a few hours of calcination at 800 C, the samples were heated 1422$^o$ C, (close to the melting point of the $WO_3$), followed by a slow cooling at a rate of 35 $^o$ C/hr to 1300 $^o$ C. Finally the tubes were `furnace-cooled` to ambient temperature. The yellow-gray powders, were analyzed by powder X-ray diffraction (XRD) studies which confirmed the purity of the products. XRD measurements on $W_{0.975}Nb_{0.025}O_{3-\delta}$ at 120 and 300K have been performed with a Enraf Nonius diffractometer using a MoKa radiation. The unreacted starting material $Nb_2O_5$+$WO_3$ mixture has the orthorhombic structure, (Fig.1, bottom) in accordance with Ref.3 ( $Nb_2O_5$ is not detected in the XRD), and the amount of $Nb_2O_5$ refers to nominal concentrations.

The dc magnetic measurements on polycrystalline powders in the range of 5-300 K were performed in a commercial (Quantum Design) superconducting quantum interference device magnetometer (SQUID). The magnetization was measured by two different procedures. (a) The sample was zero-field-cooled (ZFC) to 5 K, a field was applied and the magnetization was measured as a function of temperature. (b) The sample was field-cooled (FC) from above 300K to 5 K and the magnetization was measured.

**Experimental results.**

ZFC and FC magnetic measurements on $W_{0.99}Nb_{0.01}O_{3-\delta}$ were performed over a broad range of applied magnetic fields, and typical paramagnetic $\chi(T)$ =M/H curves measured at 10 kOe. is shown in Fig.2. The two ZFC and FC branches coincide at all temperature range regardless of the applied field, and no sign for a diamagnetic signal (which is the conventional signature of superconductivity) is observed. The M(H) curve at 5 K is linear up to 50 kOe. The susceptibility of $W_{0.99}Nb_{0.01}O_{3-\delta}$ adheres perfectly to the Curie-Weirs (CW) law: $\chi(T)=\chi_0+C/(T-\theta)$, where $\chi_0$ is the temperature independent part of $\chi(T)$, C is the molar Curie constant, and $\theta$ is the CW temperature. The paramagnetic values extracted are independent on the temperature range of the fitting. The values obtained are: $\chi_0$ =0.00055(1), and C=0.0063(1) emu K/mol Oe and $\theta$=-3.8(1) K. The extracted effective paramagnetic moment is $P_{eff}$= 0.22 $\mu_B$ per mole. This value is much lower than $P_{eff}$ values reported in Ref. 5 for various oxygen deficient $Nb_2O_5$-$_{2x}$ compounds.

In contrast to the orhorhomic structure of $WO_3$ and of the unheatd materials, the XRD pattern at 300 K for $W_{0.975}Nb_{0.025}O_3$ (Fig.1, top) found to be tetragonal. Least square fit yields the lattice parameters: a= 5.279(5) Å and c=3.829(4) Å; (V=106.74 $Å^3$). This tetragonal structure was found also at 120 K, and agrees well with the high temperature phase (obtained at above 725 C) labeled as $3Nb_2O_5$: $97WO_3$ solid solution in Ref. 3. Surprisingly, Fig. 3 shows a clear ferromagnetic behavior with a magnetic transition $T_C$ around 225 K. Shown in Fig 3, is background susceptibility of the unreacted $Nb_2O_5$ +$WO_3$ mixture, which is much lower than the susceptibility of $W_{0.99}Nb_{0.01}O_3$ exhibited in Fig. 2. Due to the limited solubility of $Nb_2O_5$ in $WO_3$, it is possible that the $W_{0.975}Nb_{0.025}O_{3-\delta}$ is not homogeneous and the uprise at low temperature may be related to the presence of a small fraction of a paramagnetic phase (say $W_{0.99}Nb_{0.01}O_{3-\delta}$).

The isothermal magnetization curve (Insert of Fig.3) of $W_{0.975}Nb_{0.025}O_{3-\delta}$ at 5 K is not linear and shows tendency toward saturation at 40 kOe reaching but not reaching saturation even at 50 kOe. where the magnetic moment $M_{sat}$ is 5 emu/mol a phenomenon typical to itinerant ferromagnetic compounds [7]. A linear extrapolation of M versus 1/H to 1/H =0, yields $M_{sat}$ = 6 emu/mole or $1.07*10^{-3}\mu_B$. Above $T_C$, the M/H curve adheres closely to the CW law: The paramagnetic values extracted in the range 230<T<320 K are: $\chi_0$ =0.000210(1), C=0.000440(1) emu K/mol Oe and $\theta$=192(4) K, which means that $P_{eff}$= 0.06 $\mu_B$. Note, the high positive $\theta$ obtained (very close to $T_C$) which indicates ferromagnetic interactions. The second remarkable feature is the $P_{eff}$ for $W_{0.975}Nb_{0.025}O_{3-\delta}$ is smaller than $P_{eff}$ = 0.22 $\mu_B$ obtained for the paramagnetic $W_{0.99}Nb_{0.01}O_{3-\delta}$ sample. Moreover, the high ratio ($P_{eff}/M_{sat}$ =56) for $W_{0.975}Nb_{0.025}O_{3-\delta}$ also consistent with itinerant electron magnetism order in this material.

In summary, the results of this study illustrate that in the $WO_3$-based materials there is a wide range of interesting physical properties. Most extraordinarily, these physical properties can be drastically affected by only slight variation of concentration of the substituted ions. The diamagnetic $WO_3$ becomes paramagnetic when only 1 at % of Nb is substituted for W. Increasing the Nb content leads to a ferromagnetic transition at $T_C$=225 K. The tetragonal structure at 120 and 300 K indicates clearly am intrinsic magnetic transition. Since both $W^{6+}$ and $Nb^{5+}$ ions do not have localized electrons, it is proposed that the magnetic interactions have an itinerant nature. The experimental results support strongly this assumption. It should be noted that this $T_C$ value is much higher than the values obtained in other oxide-based (e.g. $SrRuO_3$) itinerant electron ferromagnetic systems. The present compounds reminiscence the $Ca_{1-x}La_xB_6$ system, in which doping with $La^{3+}$ at a level of 0.1 at.% leads to unexpected ferromagnetic state with a Curie temperature of 600 K [8]. In the superconducting $Na_{0.005}WO_{3-\delta}$ system [1] Na diffuses to the surface, which means that on the surface, its concentration is higher. Apparently a 1% concentration of Na (if uniform) would not show superconductivity. Here, the tetragonal structure observed for $W_{0.975}Nb_{0.025}O_{3-\delta}$ (as compared to orthorhombic $WO_3$), indicates an homogeneous system.

**Acknowledgments** The research was supported by the Israel Academy of Science and Technology and by the Klachky Foundation for Superconductivity.

Fig. 1. XRD at 300 K of the orhorhombic unreacted $WO_3$ + (2.5%) $Nb_2O_5$ mixture before heat treatment and the tetragonal (top) $W_{0.975}Nb_{0.025}O_{3-\delta}$ phase after heating.

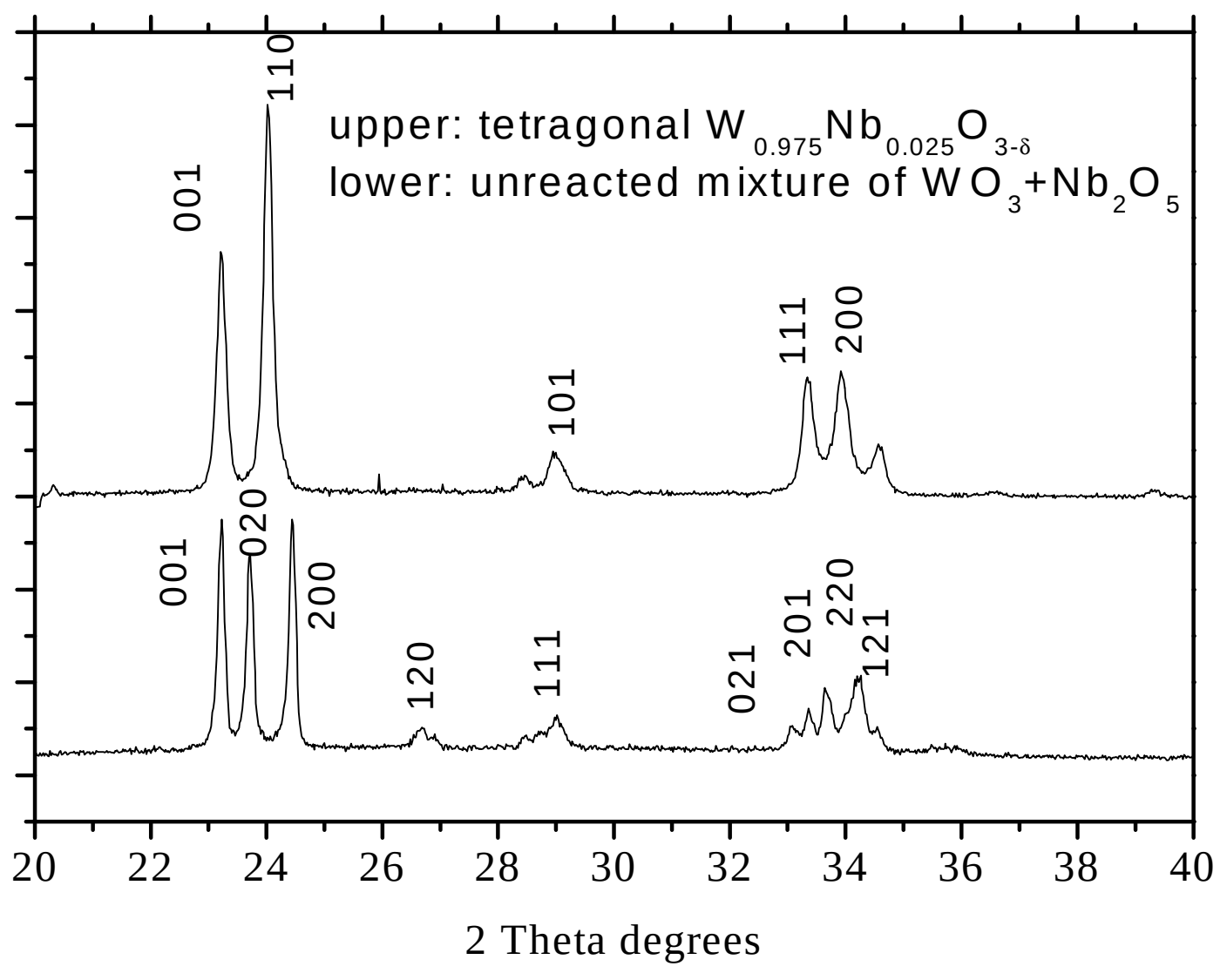

Fig. 2. M(T) of the high temperature annealed 1 mol% $Nb_2O_5$ in $WO_3$ at H=10kOe. The line is the fit to the CW law.

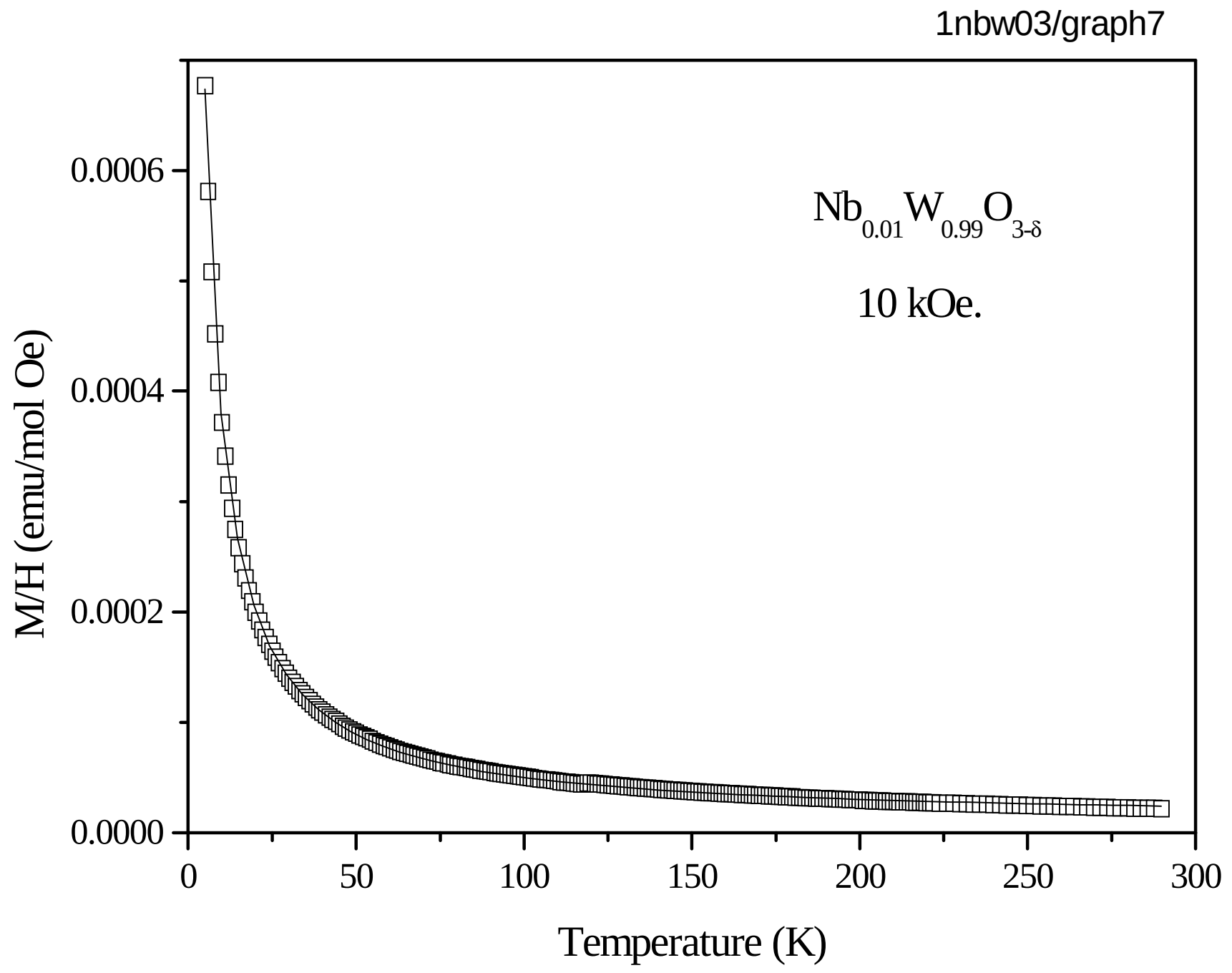

Fig. 3 ZFC and FC susceptibility curves measured at 250 Oe. and the isothermal M(H) at 5 K(inset) of itinerant ferromagnet $W_{0.975}Nb_{0.025}O_{3-\delta}$. For comparison, the susceptibility of the unreacted material is shown (bottom).

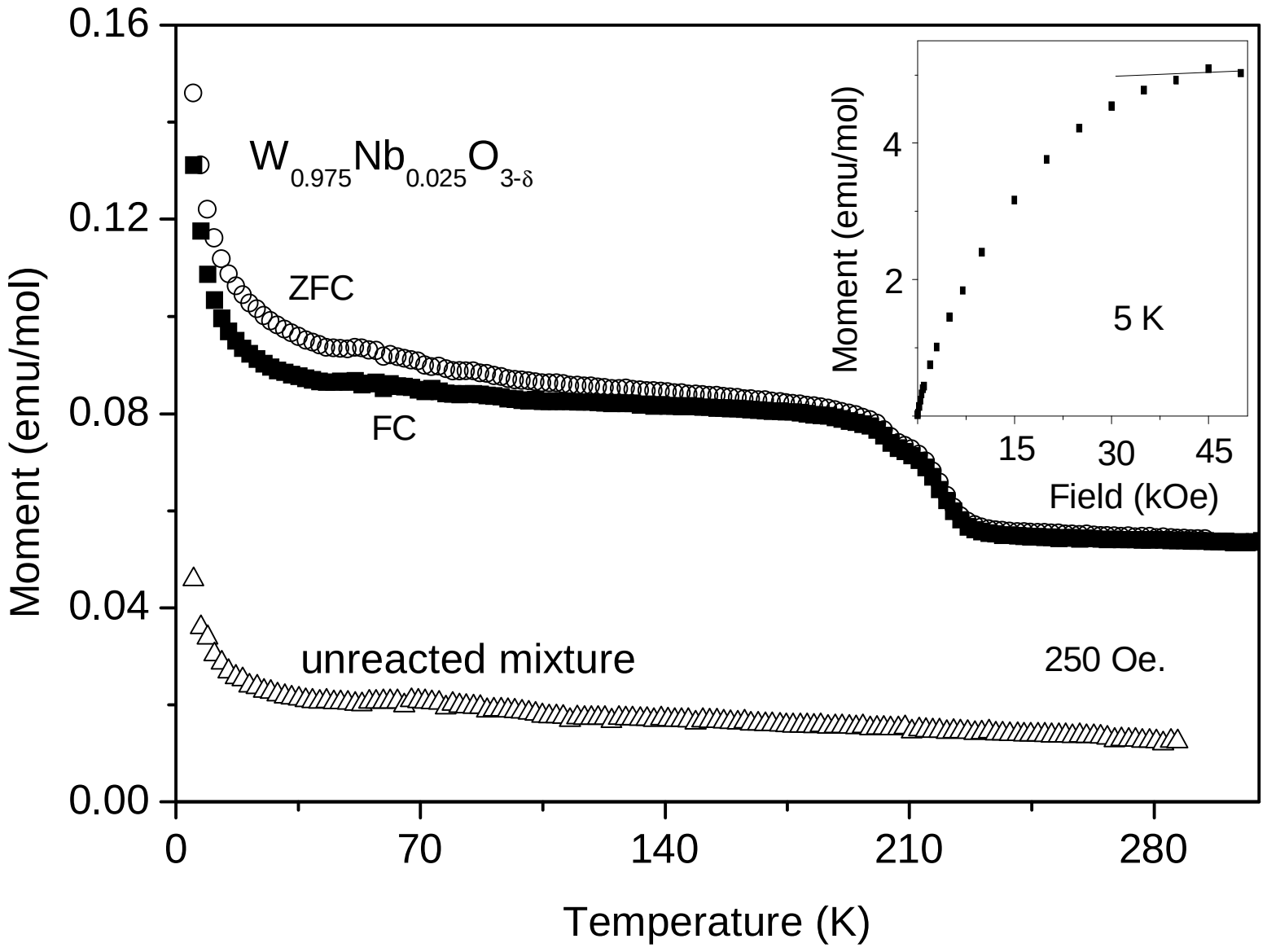